\documentclass[twocolumn,prl,showpacs,amssymb,floatfix]{revtex4}

\usepackage{graphicx}

\begin{document}

\title{\bf Negative Power Spectra in Quantum Field Theory}

\author{Jen-Tsung Hsiang}\email{cosmology@gmail.com}
\affiliation{Department of Physics, National Dong Hwa University,
Hualien, Taiwan}

\author{Chun-Hsien Wu}
\email{chunwu@scu.edu.tw}
\affiliation{Department of Physics, Soochow University \\
70 Linhsi Road, Shihlin, Taipei 111 Taiwan}

\author{L.H. Ford}
\email{ford@cosmos.phy.tufts.edu}
\affiliation{Institute of Cosmology,
Department of Physics and Astronomy \\
Tufts University, Medford, MA 02155 USA}

\begin{abstract}
We consider the spatial power spectra associated with fluctuations
of quadratic operators in field theory, such as quantum stress tensor
components. We show that the power spectrum can be negative, in
contrast to most fluctuation phenomena where the Wiener-Khinchine
theorem requires a positive power spectrum. We show why the usual
argument for positivity fails in this case, and discuss the physical
interpretation of negative power spectra. Possible applications
to cosmology are discussed.
\end{abstract}


\maketitle

The well-known  Wiener-Khinchine~\cite{Wiener,Khinchine} theorem 
states that the Fourier
transform of a correlation function is a power spectrum. 
A corollary of this theorem is that the power spectrum can normally
be written as the expectation value of a squared quantity, and
hence must be positive. The theorem is most commonly formulated
for temporal Fourier transforms. (For a review, see for example,
Ref.~\cite{Reif}.) However, it may also be formulated for spatial
Fourier transforms, which will be our focus.
The purpose of this letter is to note
that there is a loophole in the proof of the corollary, which allows
for the possibility of negative power spectra. Furthermore, we show
that such negative spectra can arise in the case of the fluctuations of
quadratic operators in quantum field theory, such as the energy
density operator.

First, let us recall the more familiar case where the power spectrum
must be non-negative. Let $F(t,{\bf x})$ be a fluctuating quantity,
e.g., a Hermitian quantum operator in flat spacetime, and let the
associated correlation function be
\begin{equation}
C({t-t',\bf x}-{\bf x'}) = \langle F(t,{\bf x}) \,F(t',{\bf x'})
\rangle\,.
\label{eq:C-x}
\end{equation}
Define $\hat{F}(t,{\bf k})$ to be a spatial Fourier transform
\begin{equation}
\hat{F}(t,{\bf k}) =  \frac{1}{(2\pi)^3}\int d^{3}x\, 
{\rm e}^{i\,\mathbf{k}\cdot\mathbf{x}} \,F(t,{\bf x}) \,,
\label{eq:F-k}
\end{equation}
and let its correlation function be
\begin{equation}
{\cal C}(t-t',{\bf k},{\bf k'}) = \langle \hat{F}(t,{\bf k}) 
\,\hat{F}({t',\bf k'})
\rangle\,.
\label{eq:C-k}
\end{equation}
If we use insert Eq.~(\ref{eq:F-k}) into Eq.~(\ref{eq:C-k}), and
change integration variables to $\mathbf{u} = {\bf x}-{\bf x'}$
and  $\mathbf{v} = {\bf x}+{\bf x'}$, we find  
\begin{equation}
{\cal C}(t-t',{\bf k},{\bf k'}) = \frac{1}{4 (2\pi)^3}\,
\delta({\bf k}-{\bf k'}) \;\int d^{3}u\, 
{\rm e}^{i\,\mathbf{k}\cdot\mathbf{u}} \, C(t-t',{\bf u})\,.
\end{equation}
Hence
\begin{equation}
{\cal C}(0,{\bf k},{\bf k'}) =  P(k)\, \delta({\bf k}-{\bf k'})\,,
\end{equation}
where the power spectrum $P(k)$ is defined by
\begin{equation}
P(k) = \frac{1}{(2\pi)^3}\int d^{3}u\, 
{\rm e}^{i\,\mathbf{k}\cdot\mathbf{u}} \,C(0,{\bf u}) \,.
\end{equation}
If ${\cal C}(0,{\bf k},{\bf k'})$ exists, then it is non-negative, 
as it is the expectation value of $\hat{F}^2(t,{\bf k})$, and as
a consequence $P(k) \geq 0$. This is a form of the Wiener-Khinchine
theorem for spatial Fourier transforms.

A possible loophole can arise if ${\cal C}(0,{\bf k},{\bf k'})$
is not well-defined. A simple example can illustrate this possibility.
Consider the operator
\begin{equation}
F(t,{\bf x}) = :\varphi^2: \,,
\end{equation}
where $\varphi$ is a massless free scalar field in four-dimensional
Minkowski spacetime. The vacuum correlation function in coordinate space
may be found from Wick's theorem to be
\begin{equation}
C(\tau,r) = \frac{1}{8 \pi^4\, (r^2 -\tau^2)^2} \,,
\end{equation}
where $\tau = t-t'$. The spatial Fourier transform is
\begin{equation}
\hat{C}(\tau,k) =  \frac{1}{(2\pi)^3}\int d^{3}x\, 
{\rm e}^{i\,\mathbf{k}\cdot\mathbf{x}} \, C(\tau,r)
= - \frac{\sin (k \tau)}{64 \pi^5\, \tau}\,.
\end{equation}
This result may be verified either by direct evaluation of the above
integral using contour integration methods, or by checking that
\begin{eqnarray}
& &C(\tau,r) = \int d^{3}k\, 
{\rm e}
^{-i\,\mathbf{k}\cdot\mathbf{x}} \, \hat{C}(\tau,k) \nonumber \\
\!\!\!\! &=&  \!\!\!-\frac{1}{16 \pi^4 r \tau} \, \lim_{\alpha \rightarrow 0}\,
\int_0^\infty  \!\!\!  dk\, k\, \sin(kr)\, \sin(k \tau) \, {\rm e}^{-\alpha k}
  \,.
\label{eq:inv-trans}
\end{eqnarray}
The associated power spectrum is
\begin{equation}
P(k) = \hat{C}(0,k) = - \frac{k}{64 \pi^5}\,, 
\label{eq:model_power}
\end{equation}
which is negative. Clearly, the positivity argument
given above does not hold in this case, and the
reason must be that ${\cal C}(0,{\bf k},{\bf k'})$ does not exist as a 
well-defined quantity.

The explicit form for $F({\bf x},t)$ in
terms of creation and annihilation operators is
\begin{eqnarray}
F(t,{\bf x}) &=& \sum_{\mathbf{q},\mathbf{q'}} 
\frac{1}{2 V \sqrt{\omega \omega'}} \left\{ 
{\rm e}^{i[(\mathbf{q}-\mathbf{q'})\cdot {\bf x} -(\omega-\omega')t]}
\; a^\dagger_{\mathbf{q'}}\, a_{\mathbf{q}} \right.\nonumber \\
&+&  \left.
{\rm e}^{i[(\mathbf{q}+\mathbf{q'})\cdot {\bf x} -(\omega+\omega')t]}
\; a_{\mathbf{q}}\, a_{\mathbf{q'}} + H.c. \right\} \,.
\end{eqnarray}
Here we use box normalization in a volume $V$,  H.c. denotes the
Hermitian conjugate, $\omega =|\mathbf{q}|$, and  $\omega'
=|\mathbf{q'}|$. The Fourier transformed operator is
\begin{eqnarray}
\hat{F}(t,{\bf k}) &=& \frac{1}{2(2\pi)^3} \, \sum_{\mathbf{q}}
\frac{1}{\sqrt{\omega \omega_1}} \, \left[ {\rm e}^{i(\omega_1 -\omega)t}
\; a^\dagger_{\mathbf{q}+\mathbf{k}} \, a_{\mathbf{q}} \right. \nonumber \\
&+&  \left. {\rm e}^{-i(\omega +\omega_1)t}
\; a_{-\mathbf{q}-\mathbf{k}} \, a_{\mathbf{q}} + H.c. \right] \,,
\end{eqnarray}
where  $\omega_1 =|\mathbf{q}+\mathbf{k}|$.

The correlation function for $\hat{F}({\bf k},t)$ can be shown to be
\begin{equation}
{\cal C}(\tau,{\bf k},{\bf k'}) = \frac{\delta_{{\bf k},{\bf k'}}}{2
  (2\pi)^6} \,  \sum_{\mathbf{q}} \frac{1}{\omega \omega_1}\,
 \cos[(\omega_1 -\omega)\tau]  \,,
\end{equation}
where $k = |{\bf k}|$.
If we take the limit $\tau \rightarrow 0$, this quantity is
formally positive, but divergent and hence undefined. This is the
positive quantity that would appear in the Wiener-Khinchine theorem
and implied a positive power spectrum had it been well-defined.
If we first take the large $V$ limit and perform the integration
on $\mathbf{q}$, then the result is
\begin{equation}
{\cal C}(\tau,{\bf k},{\bf k'}) = - \delta({\bf k}-{\bf k'})\;
\frac{\sin (k \tau)}{64 \pi^5\, \tau} \,,
\end{equation}
where we have used $V \,\delta_{{\bf k},{\bf k'}}/(2\pi)^3 \rightarrow
\delta({\bf k}-{\bf k'})$. Now we can take the $\tau \rightarrow 0$
limit, and obtain the negative power spectrum in
Eq.~(\ref{eq:model_power}).

A more physically interesting example arises in the case of the
fluctuations of quantum stress tensors, such as the energy density
of the electromagnetic field. Let $\rho(t,{\bf x})$ be the
normal-ordered energy density operator, and its vacuum correlation
function be
\begin{equation}
C_{0}(\tau,r)= 
\langle \rho(t,{\bf x}) \,\rho(t',{\bf x'})\rangle =
\frac{(\tau^{2}+3r^{2})(r^{2}+3\tau^{2})}
{\pi^{4}(r^{2}- \tau^{2})^{6}}\,,
\label{eq:em_corr}
\end{equation}
The explicit form of its spatial Fourier transform  is
\begin{equation}
\hat{C}_{0}(\tau,k) = 
-\frac{k^4 \, \sin(k\, \tau)}{960 \pi^5\, \tau} \,.
\label{eq:em_corr_k}
\end{equation}
Again, this result may be confirmed either by contour integration, or 
by checking that the inverse Fourier transform of $\hat{C}_{0}(\tau,k)$
is indeed $C_{0}(\tau,r)$, as illustrated in Eq.~(\ref{eq:inv-trans}).  
The power spectrum is
\begin{equation}
P_{0}(k) = \hat{C}_{0}(0,k) =  -\frac{k^5}{960 \pi^5}\,,
\label{eq:EM-power}
\end{equation}
which is again negative.

\begin{figure} 
\begin{center}
\includegraphics[height=5.5cm]{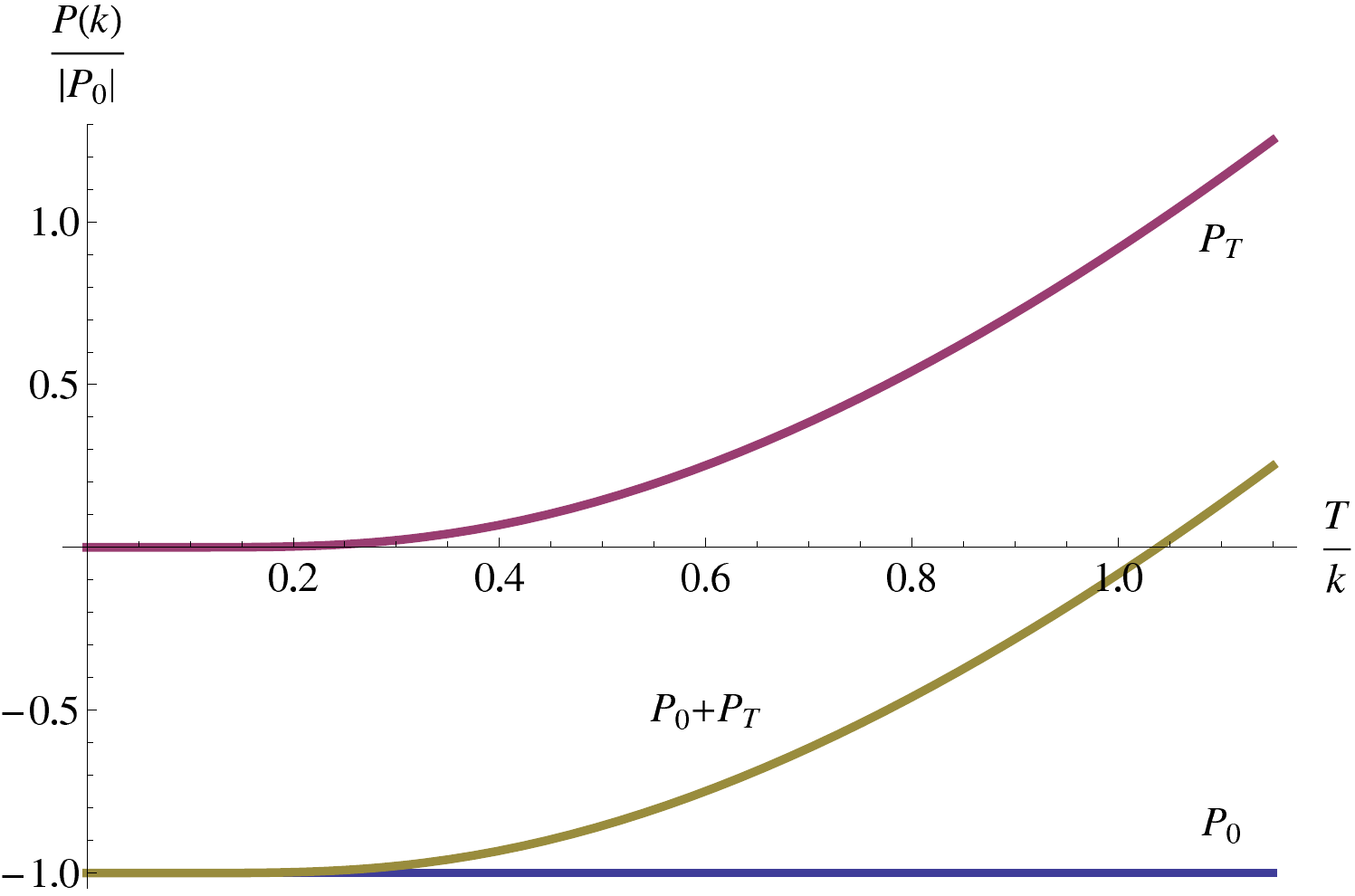}
\end{center}
\caption{The vacuum and thermal power spectra, $P_0$ and $P_T$, for the
  electromagnetic field are plotted as functions of temperature. Here
  we use
  units in which Boltzmann's constant is unity, $k_B=1$. The total
  power spectrum $P_0 +P_T$, is also plotted, and seen to be positive
for  $T > 1.04 k$.  }  
\label{fig:thermal} 
\end{figure}

This leads us to the question of the physical interpretation of
negative power spectra. One possibility is that the vacuum fluctuation
contribution is a sub-dominant contribution which reduces the
magnitude of a net positive power spectrum. This possibility can
be illustrated by energy density fluctuations at finite temperature.
The electromagnetic energy density correlation function at finite
temperature can be obtained from Eq.~(\ref{eq:em_corr}) as an image 
sum over imaginary time by replacing $\tau$ by $\tau +in\beta$,
and summing over all integers $n$. Here $\beta = 1/(k_B T)$, where
$k_B$ is Boltzmann's constant, and $T$ is the temperature. The
temperature dependent part of the correlation function can be written
as
\begin{eqnarray}
{C}_{T}(0,r) &=& 2 \sum_{n=1}^\infty {C}_{0}(in\beta,r) \nonumber \\
&=& \frac{2}{\pi^4}\, \sum_{n=1}^\infty \frac{(3r^2-n^2
  \beta^2)(r^2-3n^2 \beta^2)}{(r^2+n^2 \beta^2)^6} \,,
\end{eqnarray}
when $\Delta t=0$. The spatial Fourier transform of this expression
is $P_T(k)$, the temperature dependent part of the power spectrum,
which is found to be
\begin{equation}
P_T(k) = - \frac{k^4}{480 \pi^5 \, \beta}\; 
\ln(1 - {\rm e}^{-\beta  k})\,.
\end{equation}
Note that $P_T(k) > 0$ for all $k$, as may be seen in Fig.~\ref{fig:thermal}.
We also see from this figure that the total power, $P_0 +P_T$, 
is positive for $T > 1.04 k/k_B$.
Thus at high temperature, the negative contribution from the vacuum
fluctuations reduces the magnitude of the net power.
At lower temperatures, the vacuum term dominates, and the total
power is negative.

Consequently, we also need a physical interpretation of the case of
net negative power. Note that negative power spectra are always
associated with coordinate space correlation functions which are
singular at coincident points, and hence cannot represent the
expectation value of a meaningful squared quantity. 
However, the correlation 
function at distinct points is meaningful, and can have either sign.
If $C({t-t',\bf x}-{\bf x'}) >0$, the fluctuations at $(t,{\bf x})$
are correlated with those at  $(t',{\bf x'})$, and conversely
if  $C({t-t',\bf x}-{\bf x'}) <0$, they are anticorrelated. Changing
the sign of the power spectrum, $P(k) \rightarrow -P(k)$, changes
the sign of $C({t-t',\bf x}-{\bf x'})$, and hence interchanges
correlations and anticorrelations. So long as $C(0,0)$ is undefined,
both situations are logical possibilities. Consider the
contribution of a finite bandwidth to a spatial correlation function
and define
\begin{equation}
C_{\Delta k}(0,r) = \int_{k_0 \leq k \leq k_1}   d^{3}k\, 
{\rm e}^{-i\,\mathbf{k}\cdot\mathbf{x}} \,P(k) \,.
\end{equation}
This quantity would describe the spatial correlations in a situation
where fluctuations with $k < k_0$ or $k > k_1$ have essentially been
filtered out, It will typically be a quasi-oscillatory function in space.
Changing the sign of the power spectrum in this interval interchanges 
the minima and maxima of $C_{\Delta k}(0,r)$, interchanging correlations and
anticorrelations.

In general, it is not $C({t-t',\bf x}-{\bf x'})$ itself, but rather
integrals of the correlation function over finite spacetime regions which
are observable. Let $S_1(x)$ and  $S_2(x')$ be test functions which 
describes the effect of some measuring apparatus. The correlation function   
for the outcomes of the measurements described by $S_1$ and $S_2$
is
\begin{equation}
K = \int d^4x \, S_1(x) \int d^4x'\, S_2(x') \; C({t-t',\bf x}-{\bf x'})\,.
\end{equation}
Even though the function $C$ is singular at coincident points, it
is well-defined as a distribution, so $K$ is finite. Consider the case
of the vacuum energy density of the electromagnetic field, where
$C$ is given by Eq.~(\ref{eq:em_corr}). This correlation function
may be expressed as a total derivative~\cite{FR05}:
\begin{equation}
{C}_{0}(\tau,r)= -\frac{1}{3840 \pi^4} \, \nabla^2 \, \Box\,
 \nabla'^2 \, \Box' \; \ln^2 [(x-x')^2/\ell^2]\,,
\end{equation}
where $\ell$ is an arbitrary length. Here $ \nabla^2$ and $\Box$
denote the Laplacian and d'Alembertian operators, respectively, 
in $x$, and $ \nabla'^2$ and $\Box'$ the corresponding operators
in $x'$. The value of ${C}_{0}(\tau,r)$ is unchanged if $\ell$ changes.
We next integrate by parts in
the expression for $K$ and assume that the surface terms vanish, which
will be the case if  $S_1$ and $S_2$ vanish sufficiently rapidly at 
infinity. The result is
\begin{eqnarray}
K &=& \int d^4x \, \nabla^2 \, \Box\,S_1(x) \nonumber \\
&\times&
\int d^4x'\, \nabla'^2 \, \Box'\,S_2(x')  \; \ln^2 [(x-x')^2/\ell^2]\,.
\end{eqnarray}
This expression contains only an integrable, logarithmic singularity
and is hence finite. [An explicit example for $K$ is illustrated in
Fig.~5 in Ref.~\cite{FR05}.] The key point is that the singular nature
of the coordinate space correlation function, which always accompanies
negative power spectra, does not prevent quantities such as $K$ from
being well-defined.

Radiation pressure fluctuations  on a mirror can be computed as
integrals of a quantum stress tensor correlation 
function~\cite{WF01}, yielding a result which may also be derived by
an alternative approach based on photon number
fluctuations~\cite{Caves}. This is an illustration of how properly
defined integrals of singular correlation functions are physically
meaningful.

Quantum stress tensor fluctuations can potentially contribute to the
primordial density fluctuation spectra in inflationary 
models~\cite{WNF07,FMNWW10}. Here the observable quantities involve
time integrals of the energy density correlation function. Even though
the integrands are singular, the integrals are finite. In a model with 
a single scalar inflaton field, the contribution of electromagnetic
energy density fluctuations to the density fluctuation power spectrum  
can be written as [See Eq.~(88) of Ref.~\cite{FMNWW10}.]
\begin{equation}
P_{\delta \rho}(k) = \frac{\ell_P\, H\, S^2}{
30\, \pi2} \left(- S + \frac{4\pi\,H}{5\,k} \right)\,, 
\label{eq:power_dyn}
\end{equation}
where $\ell_P$ is the Planck length, $H$ is the Hubble parameter
during inflation, and $S$ is the scale factor change during inflation.
Note that the first term is negative. In the models discussed in 
Ref.~\cite{FMNWW10}, this term is dropped, as it corresponds to a 
delta-function term in the coordinate space correlation function, and
is hence not observable in measurements made in disjoint regions.
However, if there is any process which has the effect of filtering
this negative power spectrum, it would no longer be a delta-function
and hence could become observable. Classical, non-linear evolution
of the density perturbations is one possible filtering mechanism. 
Thus the appearance of this term illustrates the possibility of
negative power spectra in cosmology.  The appearance of a negative
power spectrum  of tensor perturbations in inflationary cosmology
will be discussed in a separate publication~\cite{WHNF}.

We have focused on fluctuations in space, and power spectra defined by
spatial Fourier transforms. As noted earlier, the Wiener-Khinchine
theorem can also be formulated in terms of temporal Fourier transforms
and an analogous power spectrum $P(\omega)$ can be defined. If the 
correlation function in time is finite in the coincidence limit, then
$P(\omega) \geq 0$. In the case of examples studied earlier, the
coincidence limit does not exist, but nonetheless one finds a 
positive power spectrum. For example, for the case of the vacuum
fluctuations of the electromagnetic energy density, the power spectrum
for temporal fluctuations is
\begin{equation}
P(\omega) = \int_{-\infty}^\infty d \tau \, {\rm e}^{-i \omega \tau}
\, C_0 (\tau,0) = \frac{\omega^7}{560 \pi^2} > 0\,.
\end{equation}
It is not known whether there are examples of negative temporal
power spectra.

In this letter, we have illustrated how the fluctuations of quadratic
operators in field theory can produce negative power spectra for
spatial fluctuations. A negative contribution can have the effect of 
decreasing a net positive spectrum. However, it is also possible for
the net spectrum to be negative, as in the case of vacuum energy
density fluctuations of the quantized electromagnetic field. Negative
power spectra have the opposite correlation-anticorrelation behavior
as does a positive spectrum with the same functional form.

\begin{acknowledgments}
We would like to thank Shun-Pei Miao, Kin-Wang Ng, Richard Woodard,
and the participants of the 14th and 15th Peyresq workshops for valuable
discussions. This work was supported in part by National Science
Foundation Grant PHY-0855360 and by the National Science Council, Taiwan, ROC under the Grant NSC99-2112-M- 031-002-MY3.

\end{acknowledgments}

\end{document}